\documentclass[twocolumn,twoside,slac_two]{revtex4}
\usepackage{graphicx}
\usepackage{fancyhdr}
\begin{document}

\title{Parsec-Scale Behavior of Blazars during High Gamma-Ray States}


\author{S. Jorstad$^{1,2}$, A. Marscher$^1$, I. Agudo$^{1,3}$, B. Harrison$^1$}
\affiliation{1. Boston U., 725 Commonwealth Ave., Boston MA, 02215, USA}
\affiliation{2. St.Petersburg State U., Universitetskiy pr. 28, Petrodvorets, 198504, St. Petersburg, Russia}
\affiliation{3. Instituto de Astrof´ısica de Andaluc´ıa, CSIC, Apartado 3004, 18080, Granada, Spain}

\begin{abstract}
We compare the $\gamma$-ray light curves of the blazars, constructed with data 
provided by the Fermi Large Area Telescope, with flux and polarization variations 
in the VLBI core and bright superluminal knots obtained via monthly monitoring 
with the Very Long Baseline Array at 43~GHz. For all blazars in the sample that 
exhibit a high $\gamma$-ray state on time scales from several weeks to several 
months, an increase of the total flux in the mm-wave core is contemporaneous with 
the $\gamma$-ray activity (more than a third of the sample). Here we present the 
results for quasars with the most extreme $\gamma$-ray behavior  
(3C~454.3, 3C~273, 3C~279, 1222+216, and 1633+382). The sources show that
in addition to the total flux intensity behavior, a maximum in 
the degree of polarization in the core or bright superluminal knot nearest 
to the core coincides with the time of a $\gamma$-ray peak to within the accuracy 
of the sampling of the radio data.  These argue in favor of location of 
many of $\gamma$-ray outbursts in blazars outside of the broad line region, 
either in the vicinity or downstream of the mm-wave VLBI core. 

\end{abstract}

\maketitle

\thispagestyle{fancy}


\section{INTRODUCTION}
The unprecedently detailed $\gamma$-ray light curves provided by the Fermi Large 
Area Telescope (LAT) show that blazars exhibit long-lasting (several months) activity
states characterized by several flares with $\gamma$-ray flux 
$>$10$^{-6}$~phot~cm$^{-2}$~s$^{-1}$, each with duration of $\sim$1-15~days and 
variability timescale as short as $\sim$1~hr, e.g., \citep{Abdo10a,Abdo10b,
MAR10,Fermi11a}. Although there is no doubt that the presence of a relativistic 
jet and $\gamma$-ray emission are tightly connected, e.g., \citep{LIST09},
the locations and mechanisms of the high energy origin are unclear and
highly debated, e.g., \citep{IVAN11b,TAV10}, 
with some evidence that different mechanisms and locations could be present  
even in a single source \citep{MAR10}.

We perform total and polarized intensity imaging of the parsec-scale jets of a sample 
of 35 $\gamma$-ray blazars obtained monthly with the Very Long Baseline Array (VLBA)
at 43~GHz \footnote{http://www.bu.edu/blazars/VLBAproject.html} at ultra-high 
resolution (0.1 milliarcseconds), starting in Summer 2008 
when the Fermi Gamma-Ray Space Telescope began to operate. We also undertake short 
campaigns of 2-week 
duration 2 times per year involving 3 VLBA epochs at 43~GHz for each campaign. 
The VLBA observations determine the flux and polarization of the millimeter-wave core and other components 
of the jet, as well as the kinematics and evolution of bright superluminal knots. 
We compare the $\gamma$-ray light curves of the blazars, constructed with data 
provided by the Fermi LAT, with flux and polarization variations 
in the VLBI core and bright superluminal knots. Here we present results of the
comparison for the quasars 3C~273, 3C~279, 3C~454.3, 1222+216, and 1633+382, which 
underwent exceptionally high $\gamma$-ray outbursts during the last three years.
\section{DATA REDUCTION}
We processed the VLBA data and created images in a manner identical to that described
in \cite{J05}. We modelled the images in terms of a small number of components with
circular Gaussian brightness distributions and determined polarization parameters of 
components using an IDL program that calculates  the mean values  within an area equal to that of 
the size established by the model fit. Figures 1-3 show the total and polarized 
intensity images of 3C 273, 3C 279, 3C 454.3, 1222+216, and 1633+382 along 
with identifications of moving knots. The core is a presumably stationary feature  
located at one end of the jet. Figure 4 shows the $\gamma$-ray light curves, 
as well as the light curves of flux and degree of polarization of the VLBI core
(light curves of the core and superluminal knots in the case of 3C~273).
The $\gamma$-ray light curves were constructed with a bin size of one week 
(one day for 3C 454.3 and 3C~273 during the high $\gamma$-ray state) using 
the P6 photon and spacecraft data and 
{\it v9r18p6-fssc-20101108} version of Science Tools provided by FSSC.

\begin{figure*}[t]
\centering
\includegraphics[width=110mm]{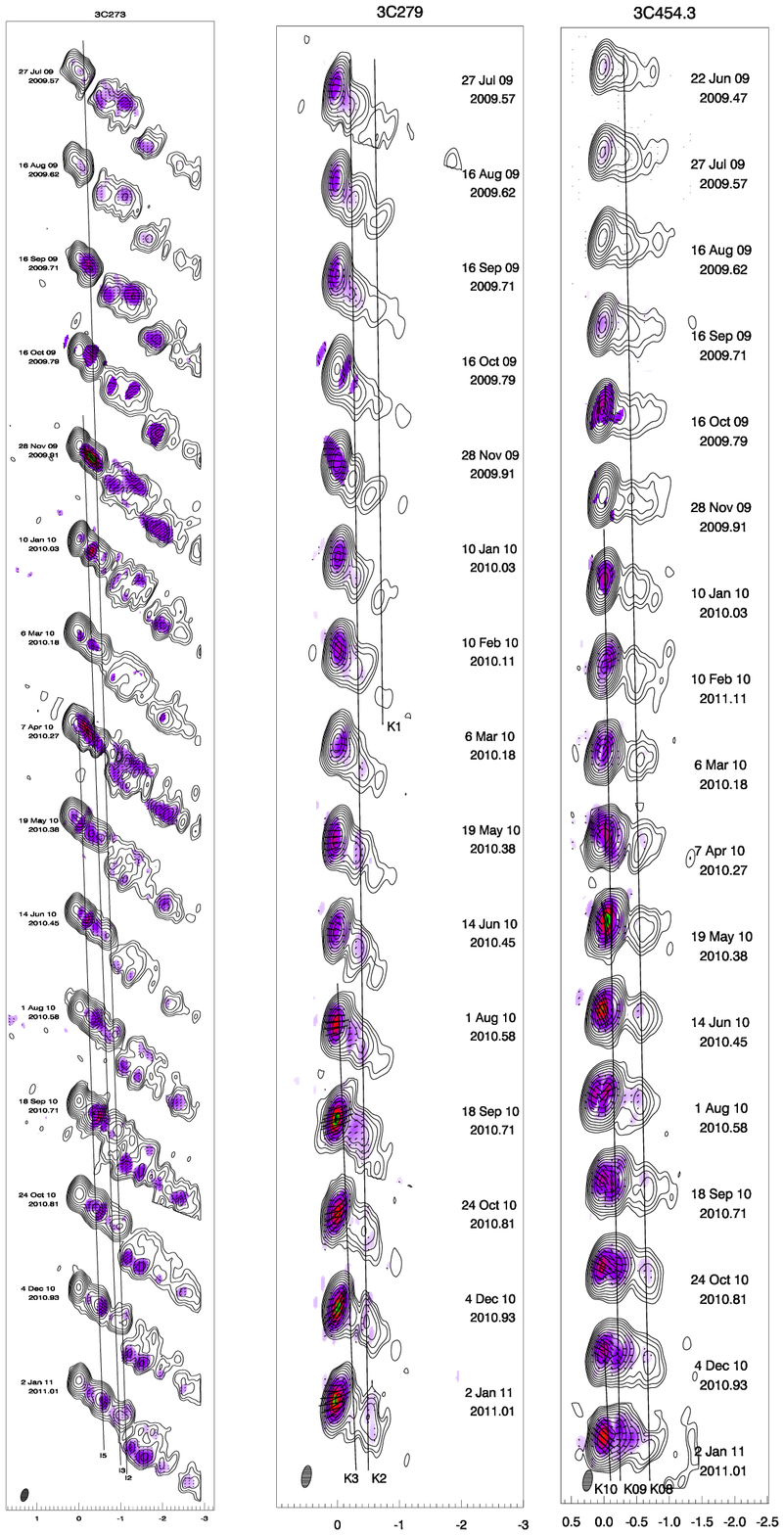}
\caption{Total (contours) and polarized (color scale) intensity images of the quasars 
3C~273, 3C~279, and 3C~454.3; black line segments show direction of the polarization;
the lowest contours are 0.25, 0.15, and 0.1\% of the total intensity peak, 
S$_{peak}$=10.0, 16.4,
and 20.3~Jy/beam, S$_{pol}$=293, 955, and 841~mJy/beam, respectively for 3C~273, 3C~279,
and 3C~454.3.} \label{vlba1}
\end{figure*}
\begin{figure*}[t]
\centering
\includegraphics[width=115mm]{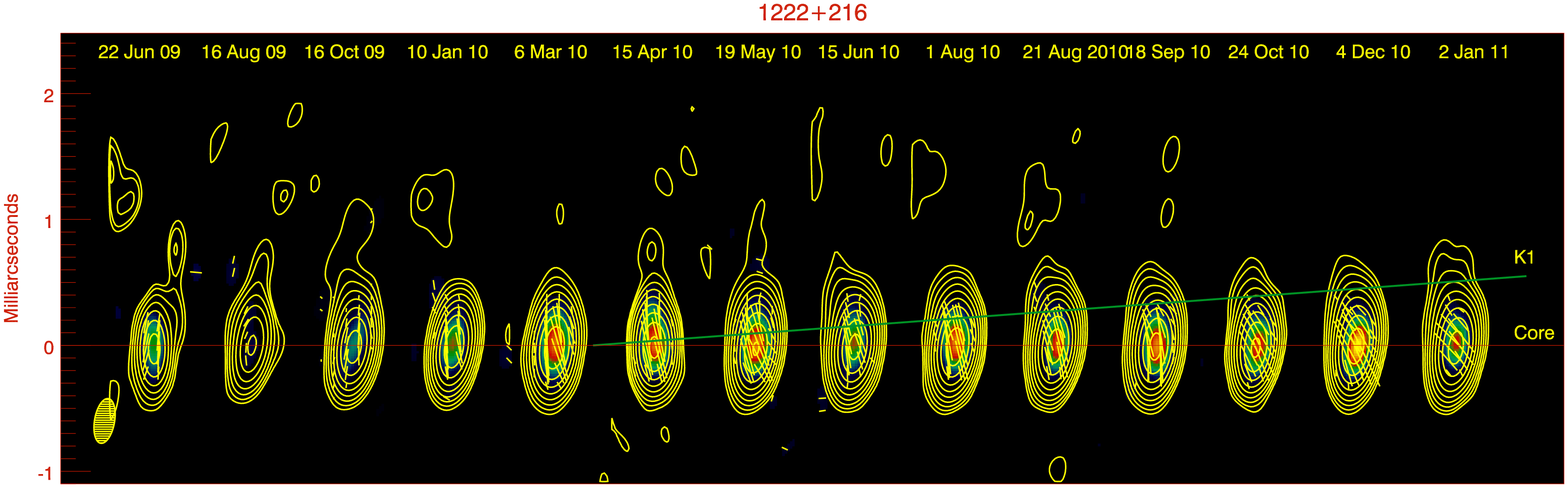}
\caption{Total (contours) and polarized (color scale) intensity images of the quasar 
1222+216; yellow line segments show direction of the polarization;
the lowest contours are 0.25\% of the total intensity peak, S$_{peak}$=1.68~Jy/beam, 
and S$_{pol}$=80~mJy/beam.} \label{vlba2}
\end{figure*}
\begin{figure*}[]
\centering
\includegraphics[width=115mm]{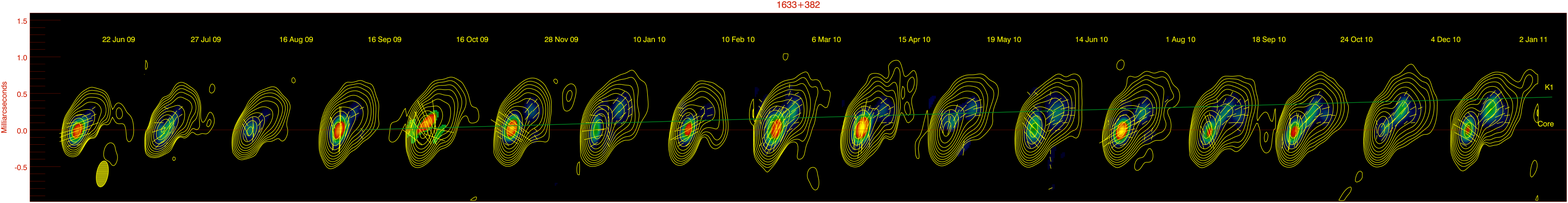}
\caption{Total (contours) and polarized (color scale) intensity images of the quasar 
1633+382; yellow line segments show direction of the polarization;
the lowest contours are 0.15\% of the total intensity peak, S$_{peak}$=2.64~Jy/beam, 
and S$_{pol}$=70~mJy/beam.} \label{vlba2}
\end{figure*}
\begin{figure*}[b]
\centering
\includegraphics[width=115mm]{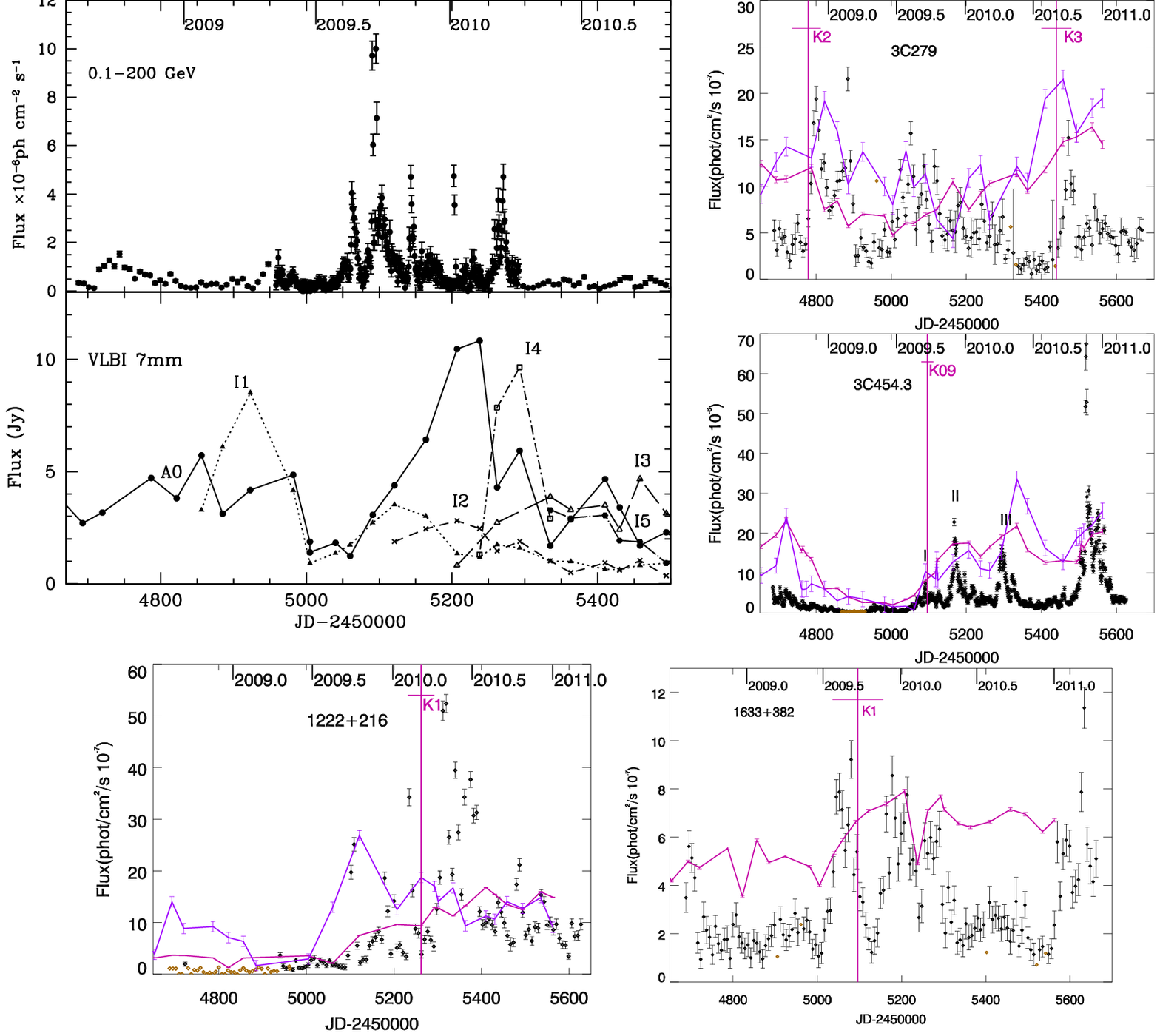}
\caption{{\it Upper Left:} $\gamma$-ray light curve (top panel) and light curves
of the VLBI components (the second panel) for the quasar 3C~273. {\it Right \& Bottom:} 
$\gamma$-ray light curves (black circles; brown circles – upper limits) of 3C~279, 3C~454.3,
1222+216, and 1633+382, total intensity light curves of the VLBI core at 43 GHz 
(pink, in Jy, multiplied by factor of 3 for 1222+216 and 16333+382), 
fractional polarization curves of the core (violet), in \%, multiplied by factor 
of 3; vertical pink lines show the time of passage of superluminal knots 
through the core.} \label{curves}
\end{figure*}
\section{DISCUSSION}
In all 5 objects, which exhibit dramatic $\gamma$-ray activity during 2008-2010, we 
observe a simultaneous increase of total intensity and fractional polarization 
in the core at 43~GHz (7~mm) accompanied by the appearance of superluminal knots in the jet.

{\bf The quasar 3C 273:}
The light curve in Figure 4 shows a high $\gamma$-ray state of the quasar during 2009.6-2010.3
(JD: 2455050-2455300) featuring 5 prominent $\gamma$-ray peaks.
We have identified 7 superluminal knots (disturbances) in the jet of 3C~273 within 
2~mas of the core (Fig. 1 \& 4), out of which 4 components (I2, I3, I4, and 
I5) appeared in the jet during the $\gamma$-ray events.
The apparent speeds  of the components range from 7c to 12c. These four components 
were ejected within 1~yr (from 2009.5 to 2010.5), while the average rate of ejection of 
superluminal knots in 3C~273, e.g., in 1998-2001, is 0.7 knots per year \citep{J05}.
This indicates a significant increase of activity in the parsec-scale jet 
contemporaneous with the $\gamma$-ray events. The fastest knot, I2 (12.0$\pm$0.7~c), 
can be associated with the most prominent $\gamma$-ray peak, 
while the brightest knot, I4, had maximum flux
coinciding with the last $\gamma$-ray peak. 

{\bf The quasar 3C 279:}
In 3C~279 we observe two knots (Fig. 1), K2 \& K3, whose appearance in the jet
is accompanied by an increase of flux and fractional polarization of the core (Fig. 4).
The knots have apparent speeds of 16.3$\pm$2.0~c and 19.7$\pm$2.0~c, respectively, and
their time of passage through the mm-wave core coincides with 
the two most prominent events in the $\gamma$-ray light curve, 
Dec. 2008 - Apr. 2009 and in Autumn 2010  (Fig. 4). 

{\bf The quasar 3C 454.3:}
In 3C~454.3 the $\gamma$-ray outburst in the end of 2009 had three prominent peaks 
(I, II, \& III, Fig. 4). Peak I coincides within 2$\pm$5 days with the passage of knot K09 
through the core. The knot is very bright (S$_{max}$=17 Jy) and moves with an apparent 
speed of 10.5$\pm$0.3c. Although the flux and degree of polarization
in the core increase during the radio/$\gamma$-ray events, the maximum
polarization occurred on 19 May 2010, $\sim$200 days after the start of 
the $\gamma$-ray activity and ejection of K09.
Note that the position angle of polarization (EVPA) of the core and K09 are almost
orthogonal (Fig. 1). The peak in polarization corresponds to 
the exit of the knot from the core region into the extended jet, which we 
associate with the last $\gamma$-ray event, III. Therefore, all three $\gamma$-ray
flares occurred while K09 was passing through the core region. This gives a size of the core 
region of $\sim$0.13~mas, equal to the sum of the sizes of the core ($\sim$0.06~mas)
and K09 ($\sim$0.08~mas) obtained by the model fit from January to May 2010.

{\bf The quasar 1222+216:}
The $\gamma$-ray flux from 4C21.35 increased at the end of September 2009 and remained high
until January 2011 (Fig. 4). The source was confidently detected at TeV energies on 
17 June 2010 (ATel \#2684). A very short time scale of $\gamma$-ray variability ($\sim$0.8~hr)
was found during this period \citep{Fermi11a}.  The source is strongly core dominated at
43~GHz. We see a doubling of the flux
and a significant increase of polarization in the core at 43 GHz during 
the high $\gamma$-ray state. A new superluminal knot, K1, moves down the jet at an apparent
speed of 14.2$\pm$0.4~c (Fig. 2), passing 
through the mm-wave core during the most prominent $\gamma$-ray peak (within the 1$\sigma$
uncertainty in ejection time). We also find a statistically significant correlation 
between $\gamma$-ray and optical variations as well. In addition, the 
optical linear polarization increased from $<$1\% to 6.5\% during the high $\gamma$-ray state 
while the optical EVPA rotated by 200$^\circ$  \citep{J10}. 
This behavior has a similar pattern as reported previously during high-energy
outbursts in BL Lac \citep{MAR08}, PKS~1510-089 \citep{MAR10}, and 3C 279 \citep{Abdo10a}.

{\bf The quasar 1633+382:}
In 2009 September an increase in $\gamma$-ray emission by a factor 
of $\sim$10 coincided with a gradual increase in the VLBI core flux of the quasar.
A high $\gamma$-ray state persisted for $\sim$1~yr and in early
2011 a new activity cycle started (Fig. 4). The VLBA images (Fig. 3)
show moving knot $K2$, with an apparent speed of 7.5$\pm$0.4~c, that 
passed through the core around 2-10 September 2009, 
coincident with the first peak of the 2009 $\gamma$-ray outburst. 
Comparison between the optical EVPA and that in the VLBI core
supports the idea that in the quasar 1633+382 a high $\gamma$-ray state is 
connected with processes originating near the mm-VLBI core \citep{J11}.
\section{Conclusions} 
We find that high levels of $\gamma$-ray activity in the quasars studied coincide 
with the production of new superluminal knots and their passage trough stationary 
bright features in the jet, usually the mm-wave VLBI core. Therefore, in many blazars 
$\gamma$-ray outbursts occur parsecs downstream of the central engine, at or beyond 
the mm-wave core. This agrees with the need to avoid excessive opacity to pair 
production in blazars detected at TeV energies.
\begin{acknowledgments}
This research was supported by NASA grants NNX08AV65G, NNX08AV61G,
NNX09AT99G, and NNX09AU10G, and NSF grant AST-0907893.
\end{acknowledgments}


\end{document}